\documentclass[a4paper, 10 pt, conference]{ieeeconf}
\IEEEoverridecommandlockouts
\usepackage{amsfonts}       
\usepackage{booktabs}       
\usepackage[T1]{fontenc}    
\usepackage[utf8]{inputenc} 
\usepackage{nicefrac}       
\usepackage{microtype}      
\usepackage{url}            

\usepackage{amsmath}
\usepackage{amssymb}
\usepackage{epsfig}
\usepackage{graphics}
\usepackage{graphicx}
\usepackage{lineno}
\usepackage{tikz}
\usepackage{wrapfig}
\usepackage{xcolor}

\usepackage{blkarray}
\usepackage{makecell}

\usepackage[english]{babel}
\usepackage{bm}
\usepackage[subrefformat=parens]{subcaption}

\DeclareMathOperator{\MSE}{MSE}

\title{\LARGE \bf Anomaly Detection By Autoencoder Based On Weighted Frequency Domain Loss}
\author{Masaki Nakanishi$^1$, Kazuki Sato$^2$, Hideo Terada$^1$\\\\%
    \parbox{3 in}{\centering \small $^{1}$ Open Stream, Inc.\\ Tokyo, Japan}
    \hspace{0.2 in}
    \parbox{3 in}{\centering \small $^{2}$ Nippon Telegraph And Telephone East Corporation(NTT EAST)\\Tokyo, Japan}
}
\date{May 2021}

\begin{document}

\maketitle
\thispagestyle{plain} 
\pagestyle{plain} 

\begin{abstract}
In image anomaly detection, Autoencoders are the popular methods that reconstruct the input image that might contain anomalies and output a clean image with no abnormalities. These Autoencoder-based methods usually calculate the anomaly score from the reconstruction error, the difference between the input image and the reconstructed image. On the other hand, the accuracy of the reconstruction is insufficient in many of these methods, so it leads to degraded accuracy of anomaly detection.
To improve the accuracy of the reconstruction, we consider defining loss function in the frequency domain. In general, we know that natural images contain many low-frequency components and few high-frequency components. Hence, to improve the accuracy of the reconstruction of high-frequency components, we introduce a new loss function named weighted frequency domain loss(WFDL).
WFDL provides a sharper reconstructed image, which contributes to improving the accuracy of anomaly detection. In this paper, we show our method's superiority over the conventional Autoencoder methods by comparing it with AUROC on the MVTec AD dataset\cite{bergmann2019mvtec}.
\end{abstract}

\section{Introduction}
Anomaly detection is the task of identifying whether the target is normal or anomalous. In recent years, there has been a lot of research on how to achieve this task. It can be applied to various fields, such as detecting defective products in manufacturing industries, detecting abnormal behavior in human activities, and testing diseases in the medical field.

Approaches to anomaly detection can be classified into supervised and unsupervised methods. In supervised methods, both normal images without defects and abnormal images with defects are used for training. In supervised methods, both normal images with no defects and anomaly images with defects are used for learning. And it is common to learn to classify them into normal and anomalous classes. However, in the actual application of anomaly detection, it is often not possible to collect enough anomaly images for training.However, it is impossible to collect enough images of abnormalities for learning in the field where anomaly detection is introduced in many cases. Also, there are cases where there are countless anomaly patterns or where the state of the anomaly is unpredictable, making it difficult to learn the features of the anomaly properly. For these reasons, supervised methods are employed only in a limited number of cases. For these reasons, supervised methods are applied only in limited cases. On the other hand, in unsupervised methods, only normal images are used for training, and the model learns their distributions. Then, anomaly detection is achieved using the measure of how far the target sample is from the distribution. Since it does not require anomalous images for training, there has been much research on unsupervised methods in recent years.

In this paper, we focus on anomaly detection based on Autoencoders \cite{pimentel2014review}, which is a popular unsupervised method. In this method, the normal image set is used as input, and the model is trained to be an identity mapping. The network includes the structure of the bottleneck. The bottleneck forces the model to learn a compressed representation of the training data expected to regularize the reconstruction towards the normal class. Anomaly images are not used for training the model. Thus, when an anomaly image is an input to the trained model, a normal image close to the input is reconstructed. Based on this, many methods formulate the anomaly score based on the reconstruction error \cite{bergmann2018improving, pol2019anomaly}. On the other hand, when using the Mean Squared Error (MSE) to train an Autoencoder, outputs tend to produce blurry output images that high-frequency components are lost. As a result, fine textures and edges of normal images are detected as defects, which reduces the accuracy of anomaly detection.

Various methods have been proposed for this problem. The first is the Generative Adversarial Networks (GANs). GANs are capable of reconstructing high-quality images \cite{radford2015unsupervised}, and many variations of GANs have been applied to anomaly detection \cite{sabokrou2018adversarially, schlegl2017unsupervised, schlegl2019f,akccay2019skip}. While GANs are able to generate high-quality images, they have many problems, such as learning instability and the possibility of mode collapse \cite{goodfellow2014generative}. Another Autoencoder-based approach is to use Structural SIMilarity (SSIM)\cite{wang2004image} instead of Mean Square Error for the loss function\cite{bergmann2018improving}. SSIM is known to be a measure of image similarity that is relatively close to human perception, and adopting it as a loss function has yielded better results than methods that use mean square error. However, the formulation of SSIM is parametric, and its hyperparameters need to be determined by the validation data set. On the other hand, due to the characteristics of the anomaly detection problem, it is rare that sufficient verification data will be available, so it is necessary to improve the quality of the reconstruction through non-parametric formulations.

In view of the above, we formulate a new non-parametric loss function in the frequency domain for anomaly detection based on Autoencoders. This improves the quality of the reconstruction, especially the high frequency components.

We use the Discrete Fourier Transform (DFT) to formulate the loss function in the frequency domain. The input image and the reconstructed image are transformed into the frequency domain by DFT, respectively. This is divided into an amplitude spectrum and a phase spectrum, so we consider the distance based on both. It is a known property that the L1 distance does not overestimate large values such as outliers compared to the L2 distance\cite{zhao2016loss}. By applying this method, the low-frequency components that are often included in images are not overly taken into account in the learning process. Our main contributions are summarized as follows:

\begin{enumerate}
    \item We have proposed a new Autoencoder-based method that considers the frequency domain in the problem of anomaly detection.
    \item The proposed method improves the reconstruction quality of high-frequency components by setting a weight matrix for the loss function.
    \item Compared to the existing Autoencoder-based methods, the proposed method achieved the best accuracy in anomaly detection on average.
\end{enumerate}

\section{Related Works}
\subsection{Anomaly detection based on generative models}
Various methods have been proposed for the problem of anomaly detection. Method based on Autoencoder is one of the popular methods which provide information that can be used to determine whether a pixel is a normal class or not to solve this problem.

Conventional Autoencoder methods make anomaly detection based on the difference between the input and reconstructed images when using the MSE. Since the model learns normal features from the training images, it is assumed that the missing regions will be reconstructed as normal with high quality when the quality of the reconstruction is especially low. Autoencoder methods usually create a model with a bottleneck structure and train it to be a constant function based on MSE:

\begin{equation}
    \MSE(\bm{f}, \bm{\hat{f}}) = \sum^{M-1}_{x=0}\sum^{N-1}_{y=0} (\bm{f}(x,y) - \bm{\hat{f}}(x, y))^2
\end{equation}
where $\bm{f}(x, y)$ is the intensity value of image $\bm{f}$ at the pixel $(x, y)$, $\bm{\hat{f}}(x, y)$ is the intensity value of reconstructed image $\bm{\hat{f}}$ at the pixel $(x, y)$.

This loss has the disadvantage of contributing to blurry reconstructions with lack of high-frequency components, and has the problem of increasing residuals, even though the image is normal. Therefore, it was not possible to distinguish the reconstruction error in the normal region from that in the abnormal region, and the accuracy of the abnormality determination could not be sufficiently obtained. To improve this problem, a method that incorporates SSIM into the reconstruction error has been proposed so far \cite{bergmann2018improving}. However, since the accuracy of Autoencoders by SSIM depends on hyperparameters, it is not a practical model when sufficient validation data is not available.

It is also believed that GANs can acquire the distribution of normal images and produce sharper images compared to Autoencoders \cite{radford2015unsupervised, sabokrou2018adversarially, schlegl2017unsupervised, schlegl2019f, akccay2019skip}.  However, GANs have many problems, for example, mode collapse problems, unstable learning, etc. In addition, some GAN-based methods require the search of latent space vectors (to generate the image closest to the input image) during inference, which is computationally expensive.

\subsection{Neural networks in the frequency domain}
In recent years, frequency-domain models have been proposed in the field of neural networks. The typical examples are works on super-resolution tasks \cite{fritsche2019frequency, sims2020frequency}. In \cite{fritsche2019frequency}, accurate super-resolution is achieved by applying different processing (neural networks) to the image's low and high frequency components. The approach in \cite{sims2020frequency} is similar, and efficient super-resolution has been achieved by defining a loss function in the frequency domain and varying the weights for each frequency.

\section{Method}
\label{sec3}
 Our proposed method is anomaly detection based on Autoencoder reconstruction. Our main focus is on the reconstruction accuracy of high frequency components. And the proposed method reconstructs the sharpness images compared to the existing Autoencoders. This makes it easier to detect fine texture anomalies and anomalies near edges, thereby improving the accuracy of anomaly detection. 

 In this chapter, we first define the frequency representation of an image using the Discrete Fourier Transform. Second, in order to improve the accuracy of the recovery of high frequency components, we define new weights in the frequency domain and define a loss function using these weights (WFDL). Finally, we define the anomaly score, which is the criterion for determining the anomaly.
\subsection{Image processing in the frequency domain}
 In this section, we define the Discrete Fourier Transform to deal with images in the frequency domain. And confirm its property. We used the dataset provided by MVTec \cite{bergmann2019mvtec} for our confirmation.

The main component in WFDL is the Discrete Fourier Transform (DFT). It is represented by equation (\ref{eq2}).

\begin{equation}
    \label{eq2}
    F(u, v) = \sum^{M-1}_{x=0} \sum^{N-1}_{y=0} f(x, y) e^{-j2\pi(\frac{ux}{M} + \frac{vy}{N})}
\end{equation}
where the image size is $M \times N$, $(x, y)$ denotes the coordinate of an image pixel in the spatial domain, $\bm{f}(x,y)$ is the intensity value of image $\bm{f}$ at the pixel $(x, y)$, $\bm{F} (u, v)$ is the frequency representation of the image $\bm{f}$, $(u, v)$ is the coordinate of the spatial frequency on the spectrum.

In addition, the exponential part of this equation can be rewritten by Euler's formula:
\begin{equation}
    e^{j\theta} = cos\theta + j sin\theta
\end{equation}
as follows.

\begin{equation}
    \label{eq4}
    e^{-i2\pi (\frac{ux}{M} + \frac{uv}{N})} = cos2\pi (\frac{ux}{M} + \frac{uy}{N}) - j sin2\pi (\frac{ux}{M} + \frac{uy}{N})
\end{equation}

As can be seen from equation (\ref{eq4}), the Fourier transform is a method of decomposing a function into a sum of a sine wave function and a cosine wave function with frequencies multiplying the fundamental frequency. The angle of the frequency is determined by the spectral position (u, v).

\begin{figure*}[h]
    \centering
    \begin{minipage}[]{0.3\linewidth}
        \centering
        \includegraphics[scale=0.3]{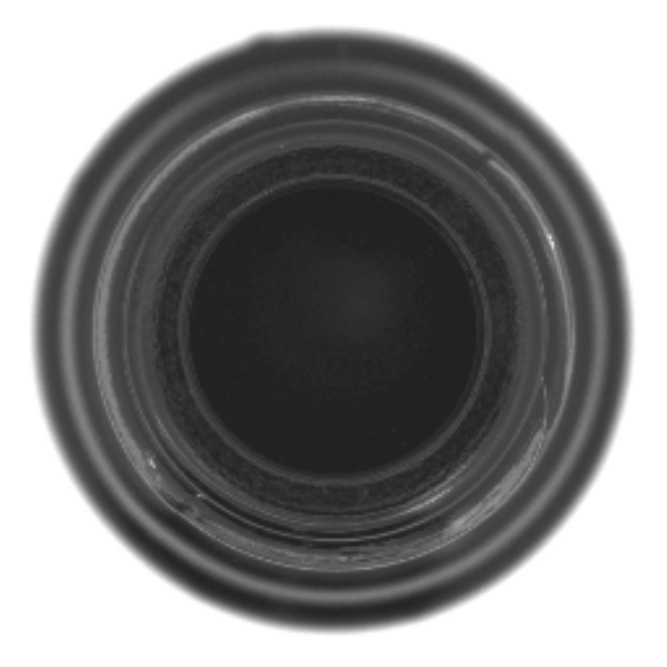}
        \includegraphics[scale=0.3]{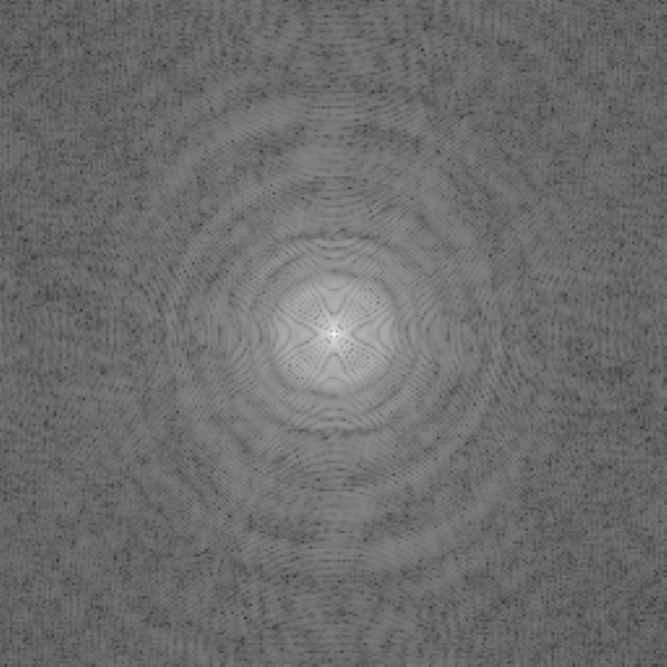}
        \subcaption{the original image}
        \label{fig:fig1a}
    \end{minipage}
    \begin{minipage}[]{0.3\linewidth}
        \centering
        \includegraphics[scale=0.3]{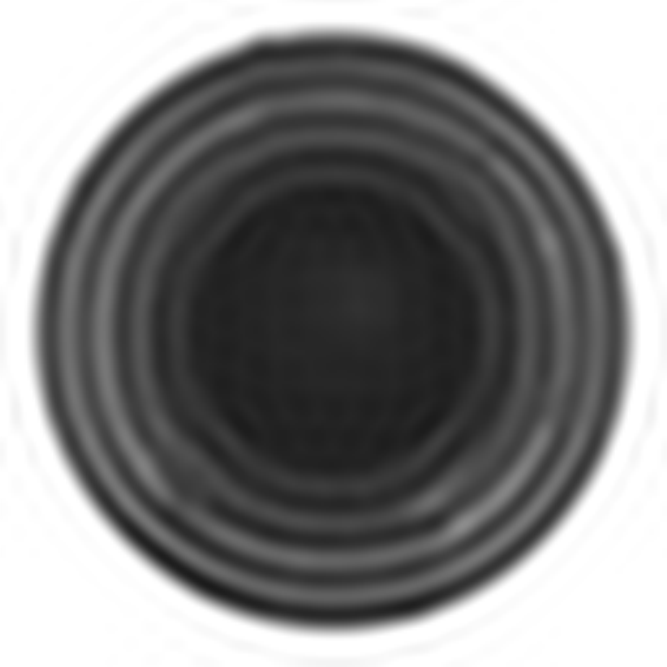}
        \includegraphics[scale=0.3]{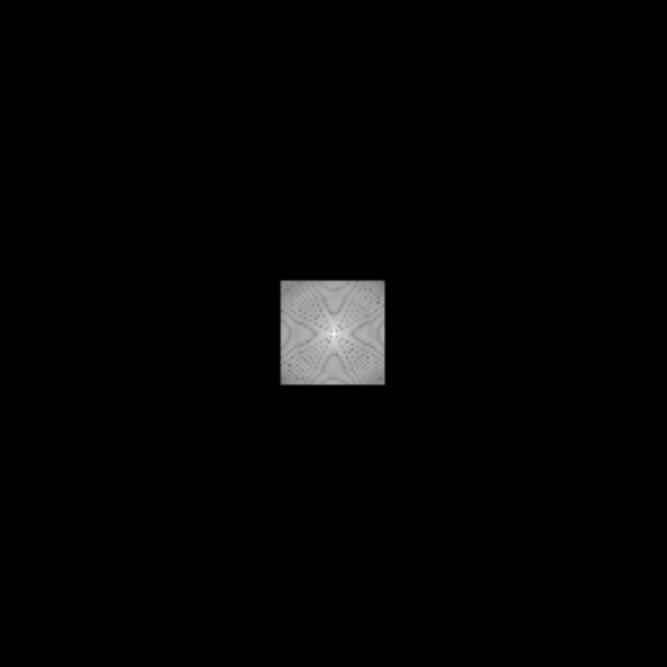}
        \subcaption{the filtered image}
        \label{fig:fig1b}
    \end{minipage}
    \begin{minipage}[]{0.3\linewidth}
        \centering
        \includegraphics[scale=0.3]{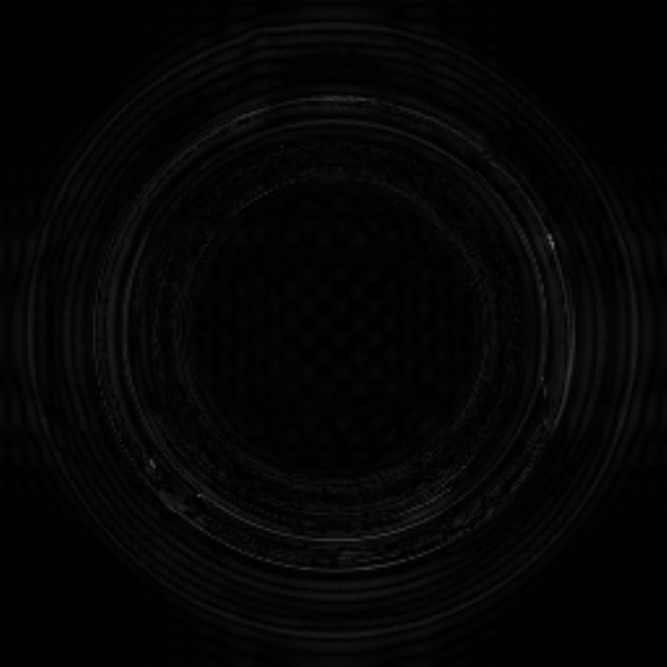}
        \includegraphics[scale=0.3]{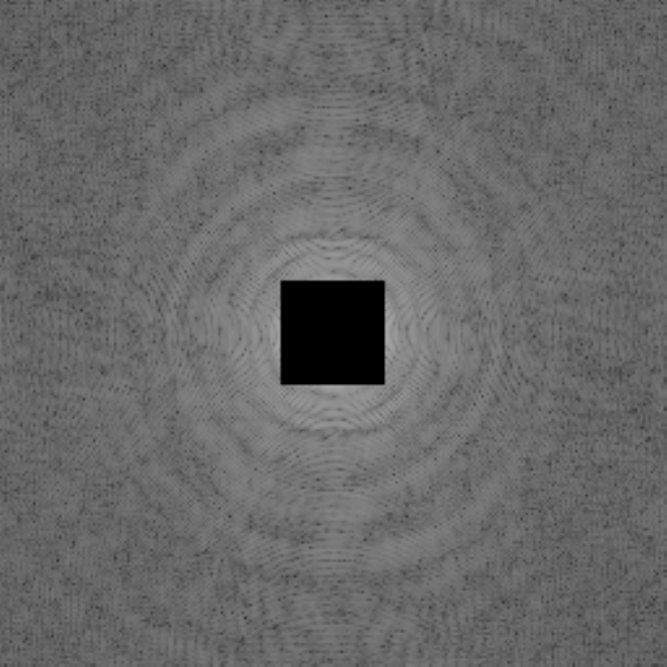}
        \subcaption{the filtered image}
        \label{fig:fig1c}
    \end{minipage}
    \caption{The result of filtering the frequency components for MVTec's bottle} \label{fig:fig1}
\end{figure*}

Fourier transform is a useful technique for frequency filtering. For a visual understanding, we show Fig.\ref{fig:fig1}. This is the result of filtering the frequency components of the MVTec bottle. In Fig. \ref{fig:fig1}, \subref{fig:fig1a} is the original image and its Fourier representation. In the Fourier representation, frequencies are lower closer to the center and higher farther away. It can be seen from \subref{fig:fig1a} that the bottle contains more low-frequency components and less high-frequency components. \subref{fig:fig1b} is the result of passing the bottle through a low-pass filter. This shows that high-frequency components such as edges and fine textures are lost. \subref{fig:fig1c} is the result of passing the bottle through a high-pass filter. This shows that low-frequency components have been lost, and only edges and fine textures remain.

\begin{figure}[]
    \centering
    \begin{minipage}[b]{0.45\linewidth}
        \centering
        \includegraphics[scale=0.3]{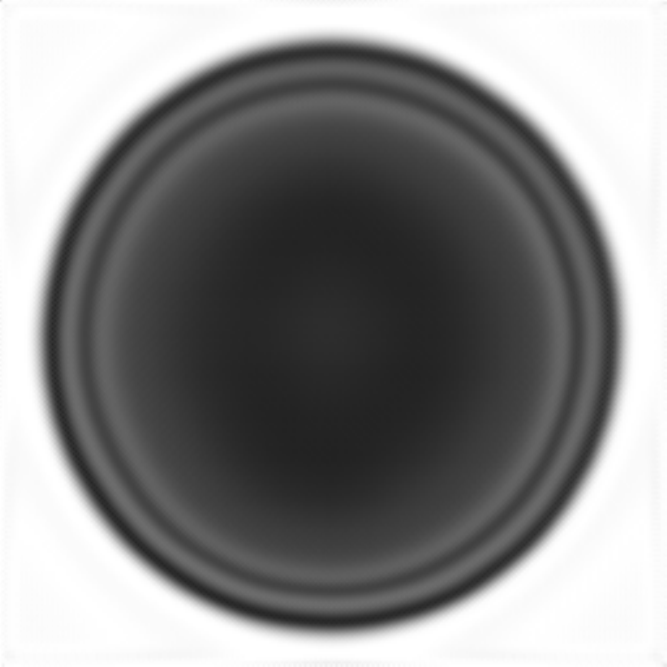}
        \subcaption{the reconstructed image}
        \label{fig:fig2a}
    \end{minipage}
    \begin{minipage}[b]{0.45\linewidth}
        \centering
        \includegraphics[scale=0.3]{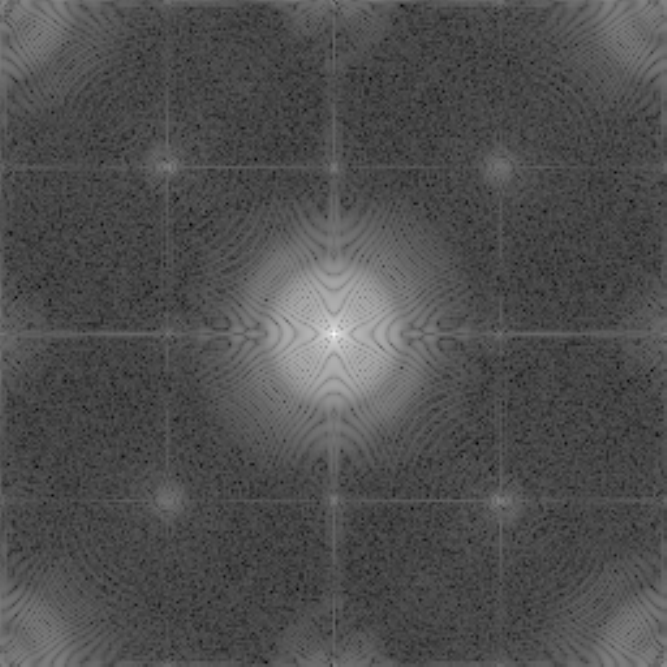}
        \subcaption{the fourier repl.}
        \label{fig:fig2b}
    \end{minipage}
    \caption{The result of the reconstruction of the bottle and corresponding Fourier representation}
    \label{fig:fig2}
\end{figure}

Now, we proceed to discuss the reconstruction images by the Autoencoder in the frequency domain. Fig. \ref{fig:fig2} shows the result of the reconstruction of the bottle by the Autoencoder based on the minimization of the L2 norm of the error and the corresponding Fourier representation. Compared to Fig. \ref{fig:fig1}\subref{fig:fig1a}, the low-frequency component is well reconstructed, but the high-frequency component is insufficient. This is because the loss function of the Autoencoder is defined as the L2 norm of the error.

In the next section, we define the loss function in the frequency domain. In addition, we propose a weighted loss function (WFDL) depending on the frequency value.

\subsection{Weighted frequency error}
To consider the loss function in the frequency domain, we define the distance between the before and after reconstructed images in the frequency domain. The image transformed into the frequency domain by the Fourier transform shown in equations (\ref{eq2}) and (\ref{eq4}) is mapped onto the complex space. Therefore, it is necessary to define a distance that considers the real and imaginary parts. Therefore, we calculate the difference in the complex plane between the images before and after the reconstruction. Then, the absolute value of the difference considering the complex numbers is calculated, and the average, equation (\ref{eq5}), is defined as the distance between images in the frequency domain.
\begin{equation}
    \label{eq5}
    d(\bm{F}, \bm{\hat{F}}) = \frac{1}{MN}\sum^{M-1}_{u=0} \sum^{N-1}_{v=0} | \bm{F}(u, v) - \bm{\hat{F}}(u, v) |
\end{equation}
\begin{equation}
    \label{eq6}
    |\bm{F}(u, v)| = \sqrt{\mathrm{Re}(F(u, v))^2 + \mathrm{Im}(F(u, v))^2}
\end{equation}
where $\mathrm{Re}()$ is a function to get the real part and $\mathrm{Im}()$ is a function to get the imaginary part.

 Furthermore, we consider weighting equation (\ref{eq5}) according to the frequency value to improve the accuracy of the reconstruction of high-frequency components. The weights should be set so that the higher the frequency, the higher values are taken. This can be expected to increase the gradient on the Autoencoder's reconstruction error of high-frequency components. When each frequency component corresponding to a two-dimensional Fourier representation is denoted by $F(u, v)$, the corresponding weight $w(u, v)$ is defined as follows:
 \begin{equation}
    \label{eq7}
    w(u, v) = \sqrt{u^2 + v^2}
 \end{equation}

Based on equations (\ref{eq5}) and (\ref{eq7}), the loss function WFDL in this paper is defined as follows:
\begin{equation}
    \label{eq8}
    L_{\mathrm{WFDL}}(\bm{F}, \hat{\bm{F}}) =  \frac{1}{MN}\sum^{M-1}_{u=0} \sum^{N-1}_{v=0} w(u, v) | \bm{F}(u, v) - \bm{\hat{F}}(u, v) |
\end{equation}

\begin{figure*}
    \centering
    \includegraphics[scale=0.3]{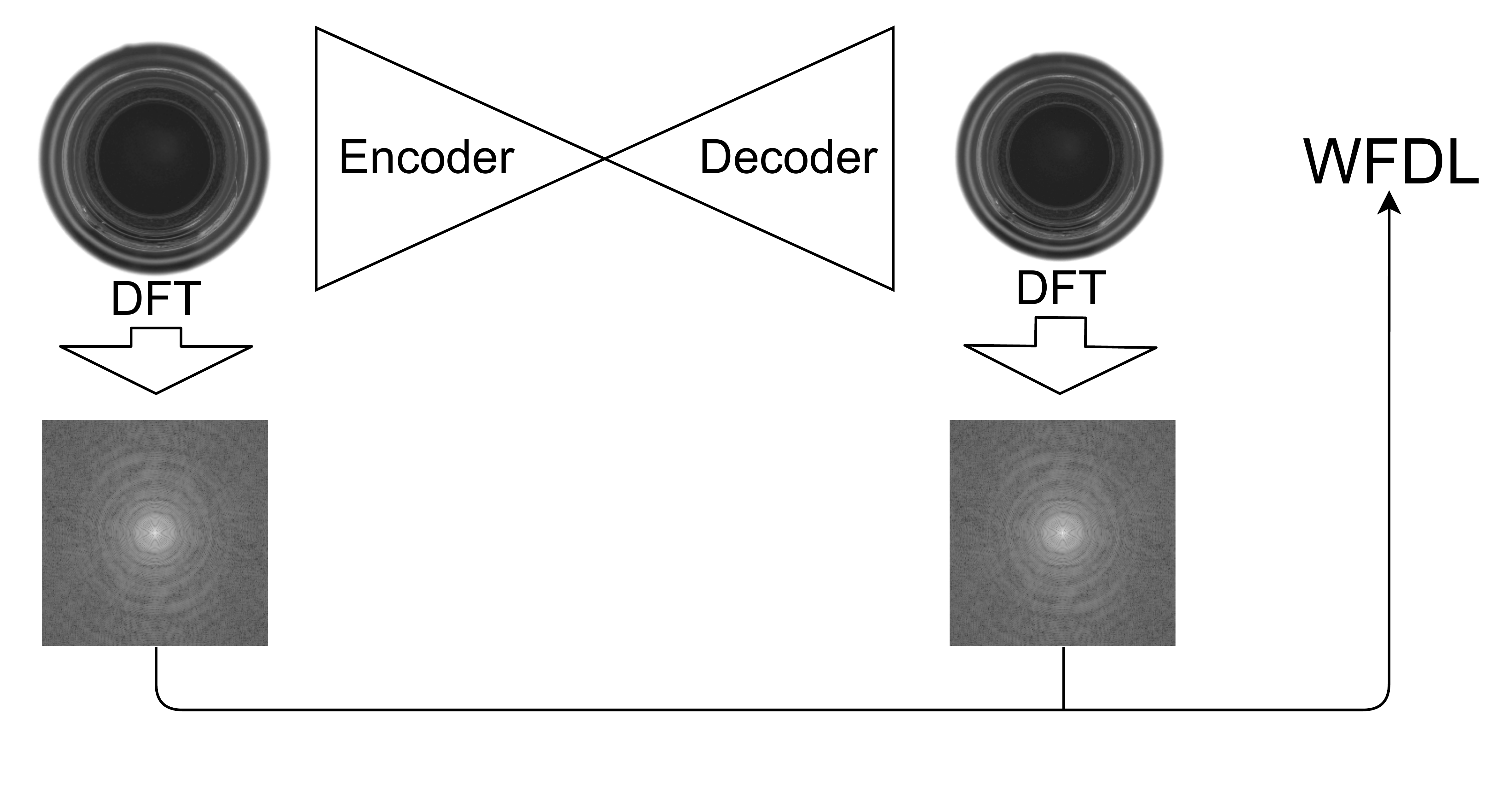}
    \caption{The architecture of the Autoencoder}
    \label{fig:fig3}
\end{figure*}

To train a model using this loss function, we first obtain a Fourier representation from the original image by 2D Fourier transform. Then, the original image is used as input to the Autoencoder to obtain the reconstructed image. Moreover, the Fourier representation of the reconstructed image is obtained by a 2-D Fourier transform as well. The Autoencoder in this paper is composed of a convolutional network. We show the architecture in Fig.\ref{fig:fig3}.

\subsection{Calculating the degree of abnormality}
Finally, to detect anomalies using the Autoencoders trained by WFDL, we define an anomaly score. Autoencoders trained by WFDL are expected to be capable of sufficiently reconstructing high-frequency components. Therefore, by calculating the anomaly score based on the difference between the input image and the reconstructed image, only the anomalous regions can be considered anomalous while edges and fine textures are not. Therefore, the anomaly score $A()$ is calculated as in equation (\ref{eq9}) based on the difference in real space, following the conventional method of Autoencoder-based anomaly detection.

\begin{equation}
    \label{eq9}
    A(\bm{f}) = \sum^{M-1}_{x=0}\sum^{N-1}_{y=0} (\bm{f}(x,y) - \bm{\hat{f}}(x, y))^2
\end{equation}

\section{Experiments}
\subsection{Experimental setup}
In this chapter, we describe the experiments to demonstrate the effectiveness of WFDL. We conducted the experiments using images from the MVTec AD dataset \cite{bergmann2019mvtec}, which contains 15 different item types. All images were resized to 256x256[px].

For all images, we trained the Autoencoders using the loss function WFDL described in Chapter \ref{sec3}. We use RAdam optimizer\cite{liu2019variance} (with an initial learning rate of 0.001, beta1 set to 0.9, beta2 set to 0.999, a weight decay set to 0.0001, and a batch size set to 64). The training ran for 2000 epochs. The architecture of the Encoder is shown in Table \ref{tab:table1}.

\begin{table}[]
    \centering
    \begin{tabular}{cc} \hline
         Block & Output Size \\ \hline
         Input & $256 \times 256 \times 3$ \\
         Residual block1 & $128 \times 128 \times 32$ \\
         Residual block2 & $64 \times 64 \times 32$ \\
         Residual block3 & $32 \times 32 \times 64$ \\
         Residual block4 & $16 \times 16 \times 64$ \\
         Residual block5 & $8 \times 8 \times 128$ \\
         Residual block6 & $4 \times 4 \times 128$ \\
         Residual block7 & $2 \times 2 \times 256$ \\
         Residual block8 & $2 \times 2 \times 256$ \\ \hline
    \end{tabular}
    \caption{The architecture of the Encoder. We use the Residual block \cite{He2016resnet}.}
    \label{tab:table1}
\end{table}

For evaluation, we use the area under the receiver operating characteristic curve (AUROC). We use the Autoencoder with L2 norm loss and the Autoencoder with SSIM loss \cite{wang2004image} for baselines.

\subsection{Results}
First, we evaluated the performance of each method quantitatively by AUROC. Table \ref{tab:table2} shows the results using the MVTec dataset\cite{bergmann2019mvtec}. This dataset has ten types of objects and five types of textures. AUROC by WFDL is the best value when averaged over these types, indicating the effectiveness of this method.

\begin{table}[]
    \centering
    \begin{tabular}{|c|c|c|c|c|}
    \hline
         \multicolumn{2}{|c|}{Category} & AE(SSIM) & AE(L2) & WFDL \\ \hline
         Texture & Carpet & $\bm{0.72}$ & $0.50$ & $0.54$ \\ \cline{2-5}
          & Grid & $0.63$ & $\bm{0.78}$ & $0.77$ \\ \cline{2-5}
          & Leather & $0.26$ & $0.44$ & $\bm{0.45}$ \\ \cline{2-5}
          & Tile & $0.33$ & $\bm{0.73}$ & $\bm{0.73}$ \\ \cline{2-5}
          & Wood & $0.86$ & $0.74$ & $\bm{0.89}$ \\ \hline
         Objects & Bottle & $0.86$ & $0.80$ & $\bm{0.94}$ \\ \cline{2-5}
          & Cable & $0.72$ & $0.56$ & $\bm{0.73}$ \\ \cline{2-5}
          & Capsule & $0.68$ & $0.62$ & $\bm{0.71}$ \\ \cline{2-5}
          & Hazelnut & $0.78$ & $0.78$ & $\bm{0.83}$ \\ \cline{2-5}
          & Metal nut & $0.62$ & $\bm{0.73}$ & $0.65$ \\ \cline{2-5}
          & Pill & $0.51$ & $0.62$ & $\bm{0.81}$ \\ \cline{2-5}
          & Screw & $0.51$ & $0.69$ & $\bm{0.82}$ \\ \cline{2-5}
          & Toothbrush & $0.92$ & $\bm{0.98}$ & $0.94$ \\ \cline{2-5}
          & Transistor & $0.78$ & $0.71$ & $\bm{0.83}$ \\ \cline{2-5}
          & Zipper & $0.57$ & $\bm{0.80}$ & $0.71$ \\ \hline
    \end{tabular}
    \caption{AUROC for anomaly detection on MVTecAD.}
    \label{tab:table2}
\end{table}

\begin{figure*}[h]
    \centering
    \begin{minipage}[]{0.18\linewidth}
        \centering
        \includegraphics[scale=0.1]{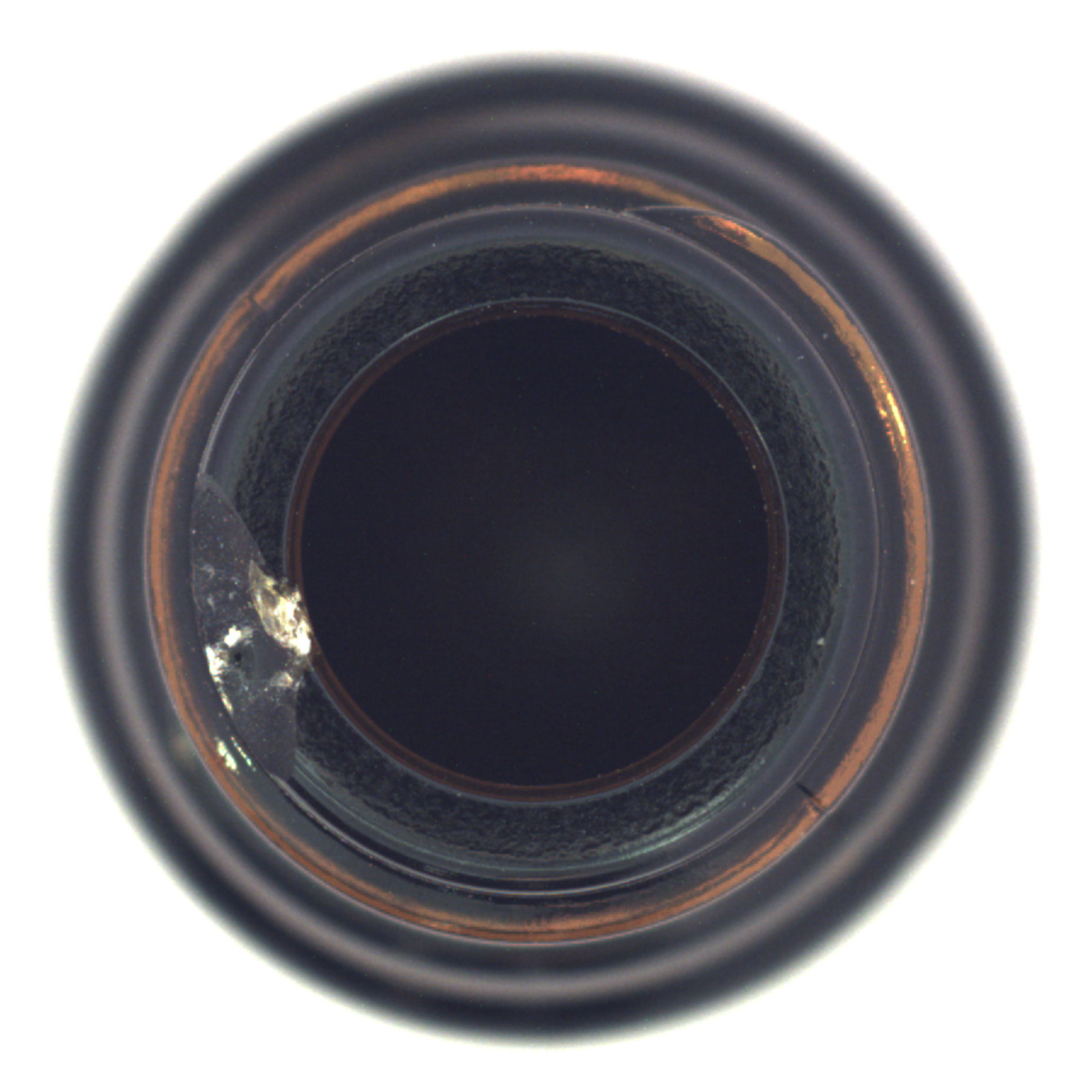}
        \subcaption{original}
        \label{fig:fig4a}
    \end{minipage}
    \begin{minipage}[]{0.18\linewidth}
        \centering
        \includegraphics[scale=0.3]{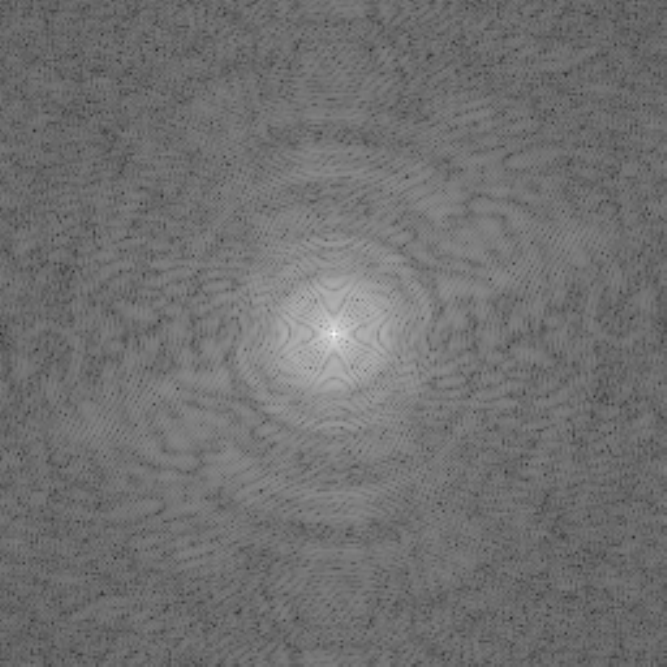}
        \subcaption{fourier repl.}
        \label{fig:fig4b}
    \end{minipage}
    \begin{minipage}[]{0.18\linewidth}
        \centering
        \includegraphics[scale=0.3]{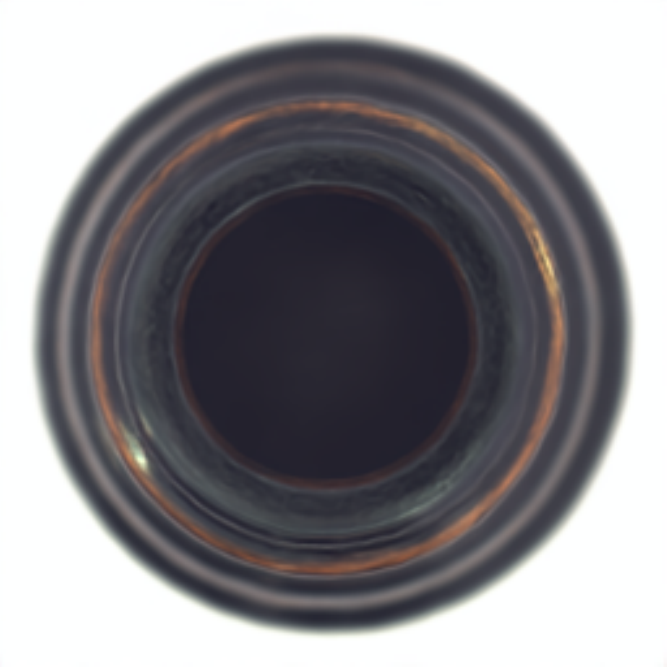}
        \includegraphics[scale=0.3]{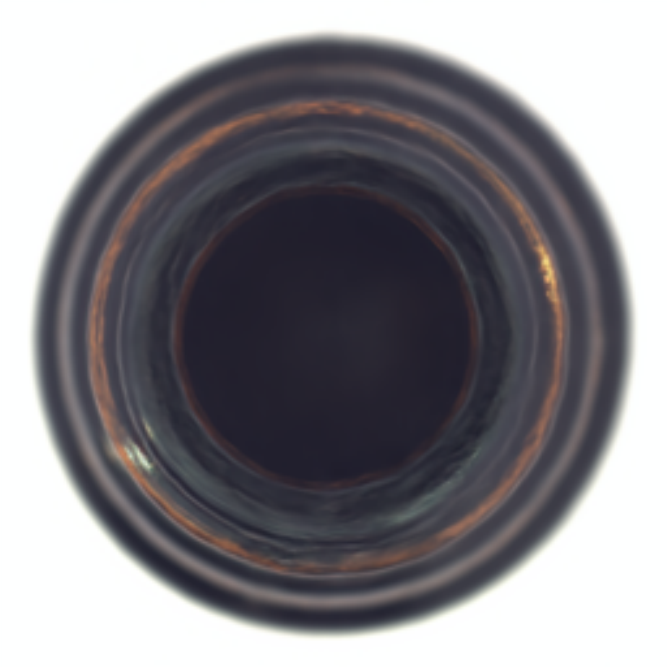}
        \subcaption{reconstructed}
        \label{fig:fig4c}
    \end{minipage}
    \begin{minipage}[]{0.18\linewidth}
        \centering
        \includegraphics[scale=0.3]{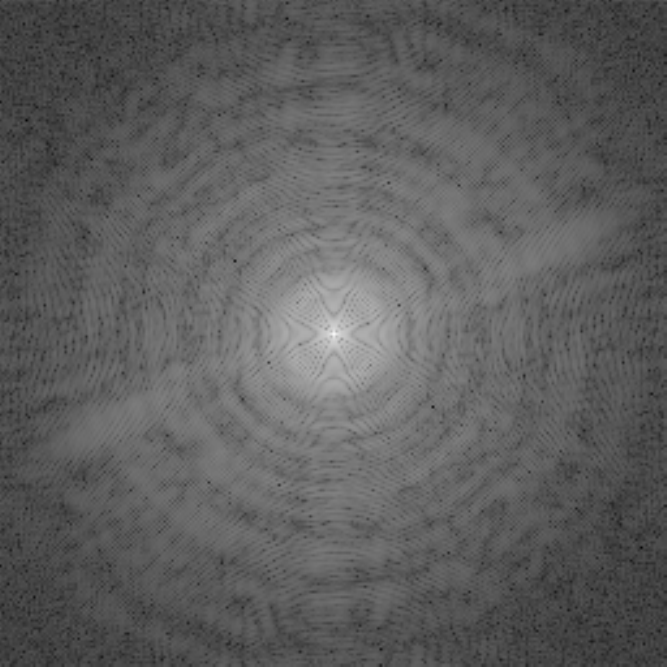}
        \includegraphics[scale=0.3]{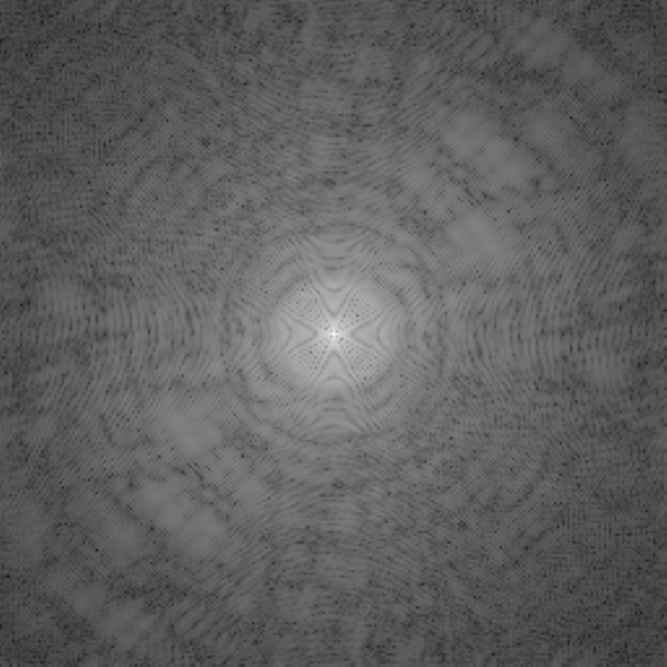}
        \subcaption{fourier repl. of the reconstructed}
        \label{fig:fig4d}
    \end{minipage}
    \begin{minipage}[]{0.18\linewidth}
        \centering
        \includegraphics[scale=0.3]{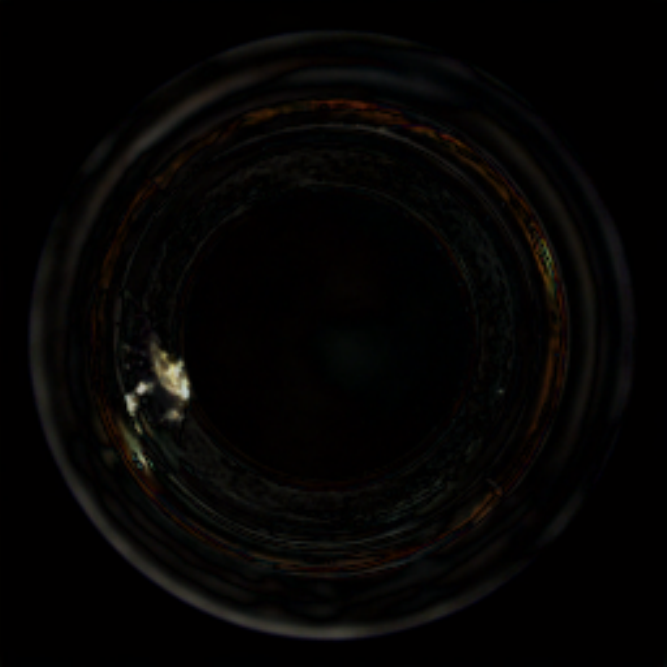}
        \includegraphics[scale=0.3]{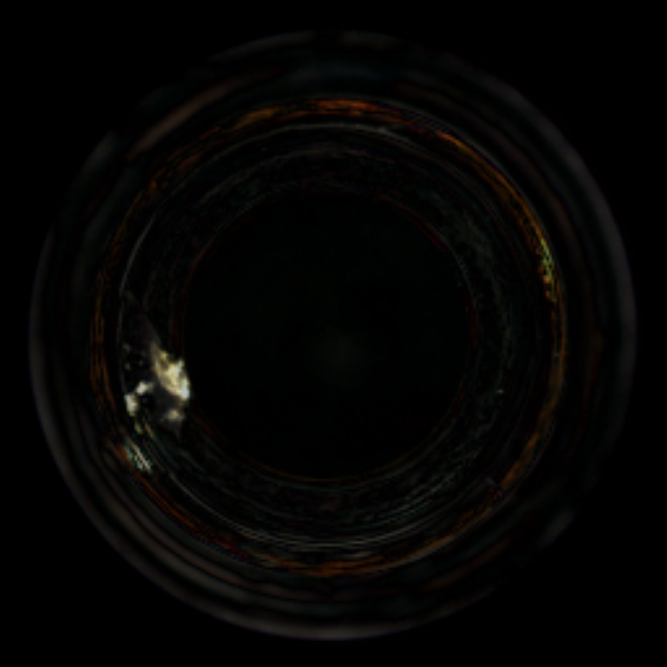}
        \subcaption{residual}
        \label{fig:fig4e}
    \end{minipage}

    \caption{The qualitative comparison of the performance, top: SSIM, donw: WFDL(bottle)} 
    \label{fig:fig4}
\end{figure*}

\begin{figure*}[h]
    \centering
    \begin{minipage}[]{0.18\linewidth}
        \centering
        \includegraphics[scale=0.1]{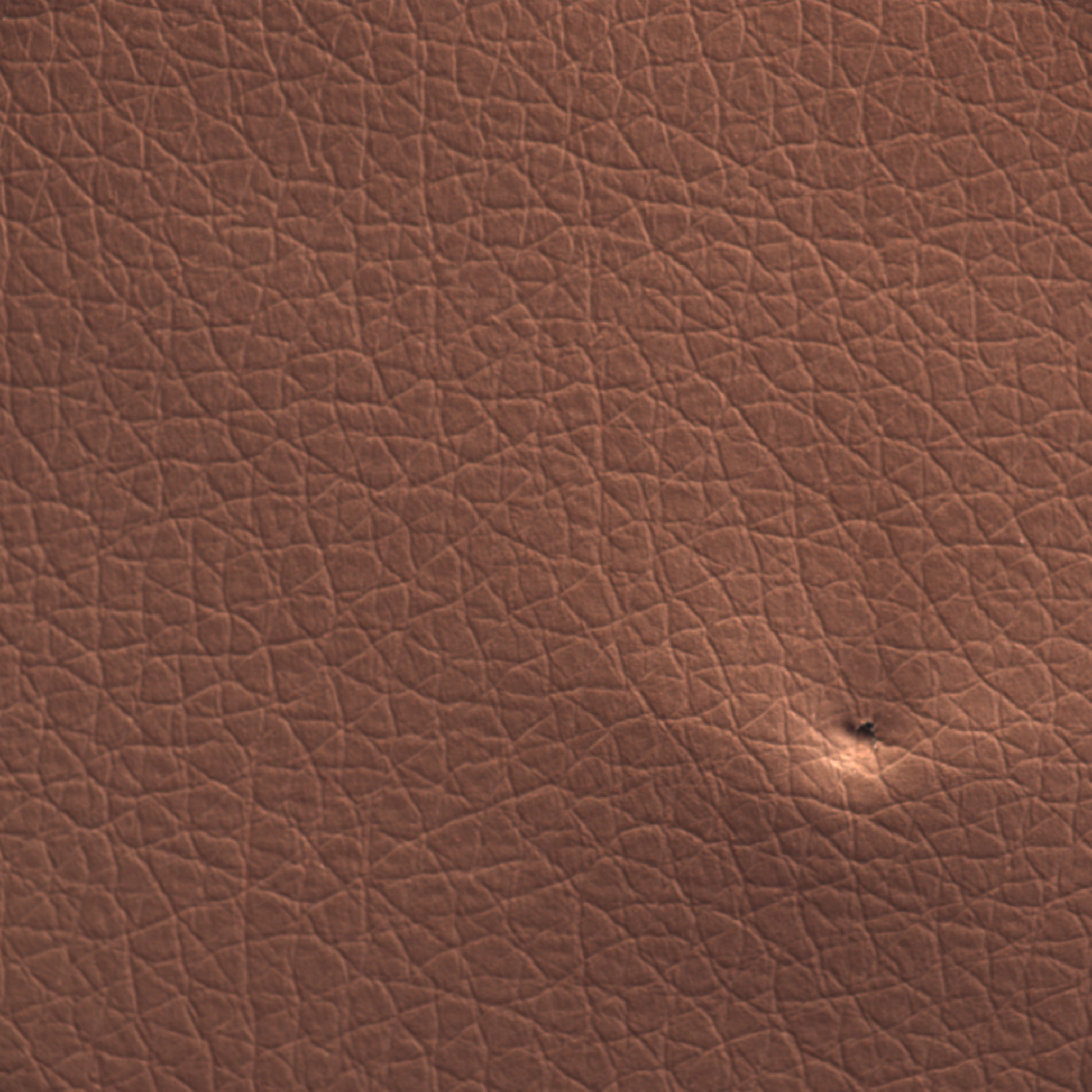}
        \subcaption{original}
        \label{fig:fig4a}
    \end{minipage}
    \begin{minipage}[]{0.18\linewidth}
        \centering
        \includegraphics[scale=0.3]{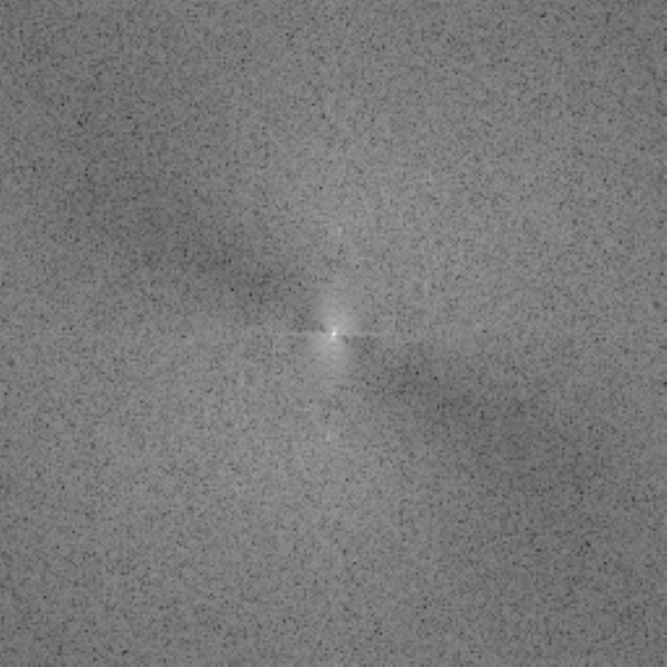}
        \subcaption{fourier repl.}
        \label{fig:fig4b}
    \end{minipage}
    \begin{minipage}[]{0.18\linewidth}
        \centering
        \includegraphics[scale=0.3]{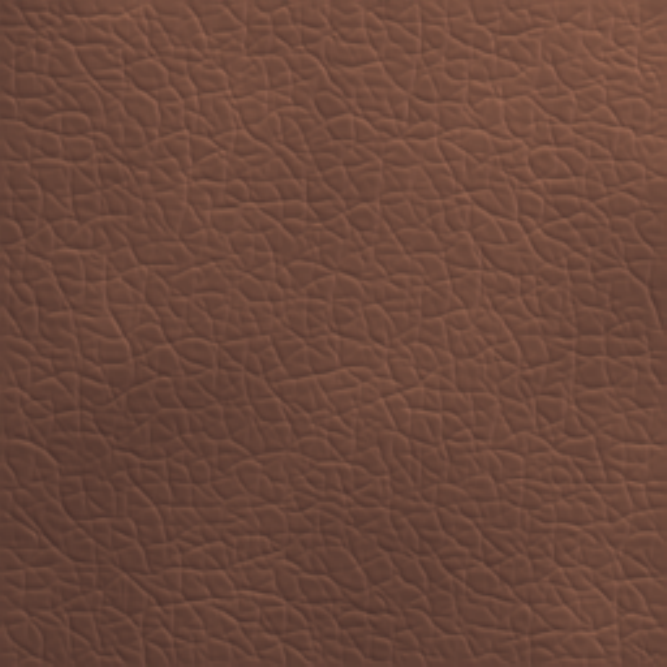}
        \includegraphics[scale=0.3]{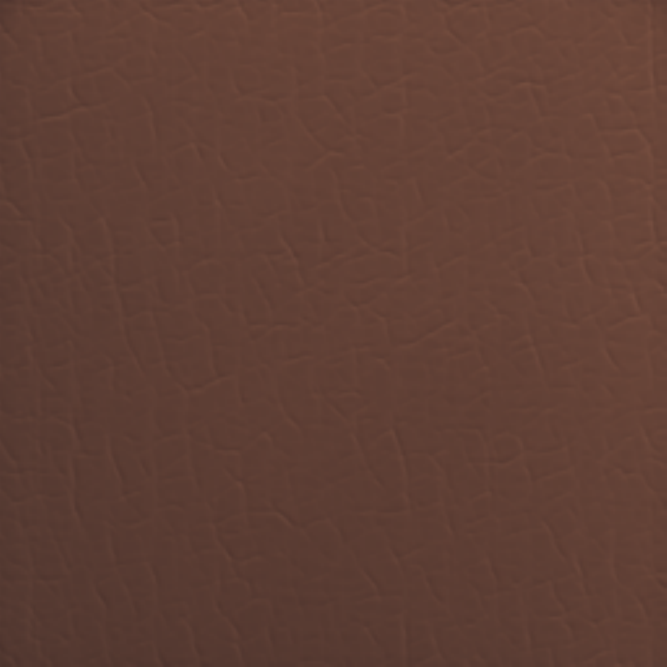}
        \subcaption{reconstructed}
        \label{fig:fig4c}
    \end{minipage}
    \begin{minipage}[]{0.18\linewidth}
        \centering
        \includegraphics[scale=0.3]{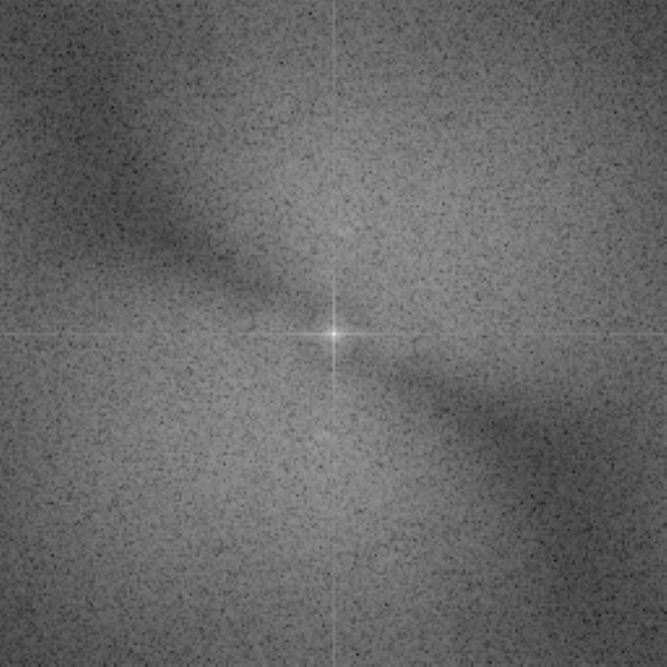}
        \includegraphics[scale=0.3]{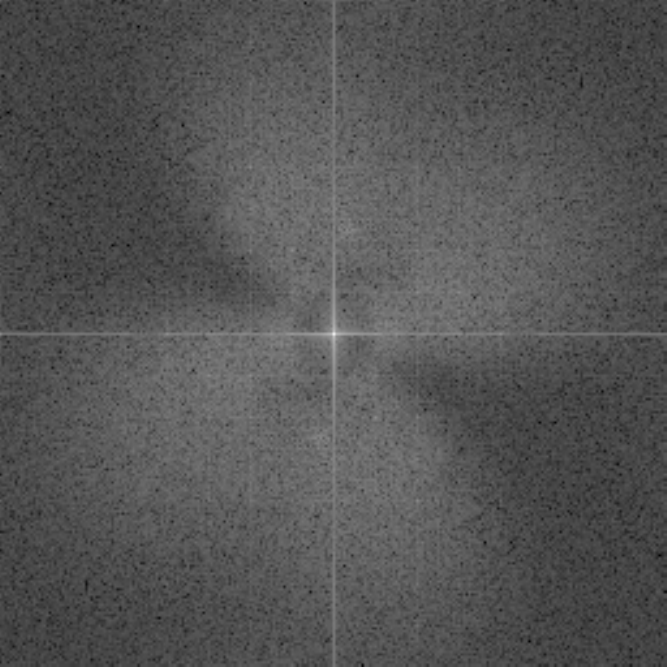}
        \subcaption{fourier repl. of the reconstructed}
        \label{fig:fig4d}
    \end{minipage}
    \begin{minipage}[]{0.18\linewidth}
        \centering
        \includegraphics[scale=0.3]{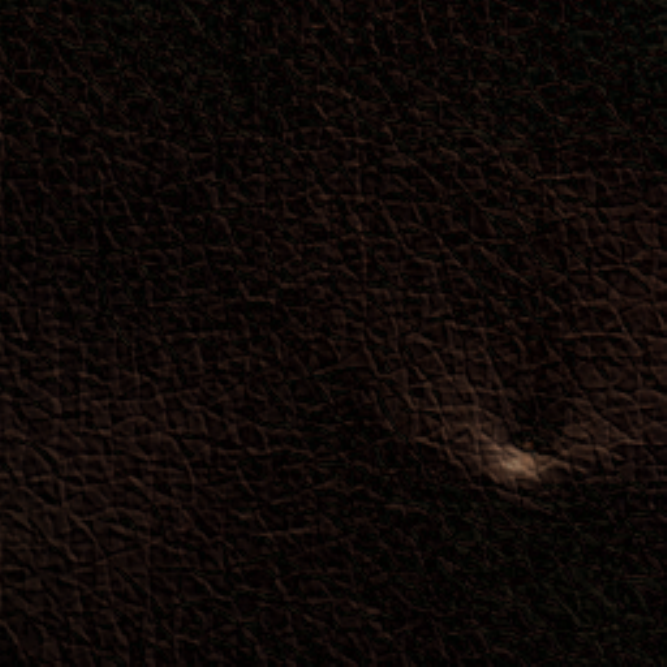}
        \includegraphics[scale=0.3]{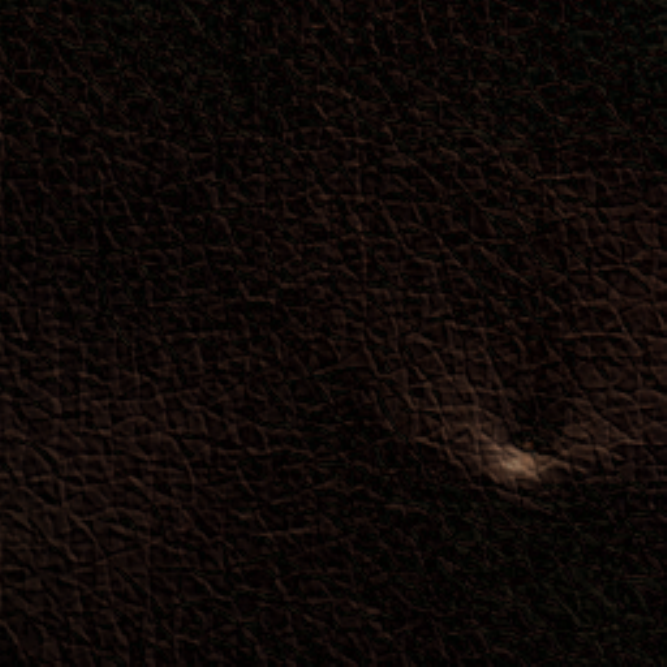}
        \subcaption{residual}
        \label{fig:fig4e}
    \end{minipage}

    \caption{The qualitative comparison of the performance(leather)} 
    \label{fig:fig5}
\end{figure*}

Fig.\ref{fig:fig4} and Fig.\ref{fig:fig5} shows a qualitative comparison of the performance for the data types bottle and leather. In Fig.\ref{fig:fig4} and \ref{fig:fig5}, (a) is the original image, (b) is the Fourier representation of the original image, (c) is the reconstructed image, (d) is the Fourier representation of the reconstructed image, and (e) is the residual image.

In the quantitative evaluation, WFDL performed relatively better than the other methods in the bottle type. In contrast, it performed rather worse in the leather type, so we will discuss these two types by checking them.

For the bottle type, the detection of defects is stable because the residuals of defective areas are large for all methods. Areas with detailed textures and areas near edges tend to have residuals even though they are not defects. However, even in these regions, the WFDL has the lowest residuals. Also, the Fourier transform of the reconstructed image shows that WFDL can reconstruct high-frequency components relatively well. In contrast, the other methods can only reconstruct low-frequency components. These observations indicate that WFDL has improved the reconstruction of high-frequency components.

On the other hand, for the leather type, the results are poor for all methods. Residuals are occurred in both defective and normal regions, making it difficult to distinguish between them. We believe that this is because the leather type contains many high-frequency components. In the WFDL reconstruction for the leather type, the reconstruction did not go well because the weight for the high-frequency component was too strong. Therefore, we believe that improvements such as dynamically changing the weight equation (\ref{eq7}) according to the distribution of the frequency components of the input image will improve the accuracy.

\section{Conclusion}
 This paper has improved the accuracy of anomaly detection based on Autoencoders by defining a new loss function (WFDL) in the frequency domain. WFDL is specifically designed to improve the quality of reconstruction of high-frequency components.  This improved the accuracy of edge and fine texture reconstruction compared to existing Autoencoder-based methods, contributing to anomaly detection accuracy. The accuracy comparison using the MVTec dataset shows that the new method gives better results than the existing Autoencoder-based methods. On the other hand, there are still issues with detecting anomalies in images that contain many high-frequency components such as leathers. However, we expect that it can improve this by adaptively modifying the weights in equation (\ref{eq7}) to the frequency components contained in the image.

\bibliographystyle{unsrt}
\bibliography{references.bib}

\end{document}